\documentclass[twocolumn,aps,prl,showpacs,superscriptaddress]{revtex4}

\usepackage{graphicx}
\usepackage{amsmath,amssymb}
\usepackage{color}

\newcommand{\be}{\begin{eqnarray}}
\newcommand{\ee}{\end{eqnarray}}

\newcommand{\beq}{\begin{equation}}
\newcommand{\eeq}{\end{equation}}
\newcommand{\beqa}{\begin{eqnarray}}
\newcommand{\eeqa}{\end{eqnarray}}
\newcommand{\ket} [1] {\vert #1 \rangle}
\newcommand{\bra} [1] {\langle #1 \vert}

\newcommand{\eins}{\mbox{$1 \hspace{-1.0mm} {\bf l}$}}

\begin{document}


\title{Experimental Tests of Classical and Quantum Dimensions}


\author{Johan Ahrens}
 \affiliation{Department of Physics, Stockholm University, S-10691,
 Stockholm, Sweden}
\author{Piotr Badzi{\c a}g}
 \affiliation{Department of Physics, Stockholm University, S-10691,
 Stockholm, Sweden}
\author{Marcin Paw{\l}owski}

\affiliation{Instytut Fizyki Teoretycznej i Astrofizyki, Uniwersytet
Gda\'{n}ski, PL-80-952 Gda\'{n}sk, Poland}
\affiliation{Department of Mathematics, University of Bristol, Bristol BS8 1TW,
United Kingdom}

\author{Marek \.Zukowski}

\affiliation{Instytut Fizyki Teoretycznej i Astrofizyki, Uniwersytet
Gda\'{n}ski, PL-80-952 Gda\'{n}sk, Poland}

\author{Mohamed Bourennane}
 \affiliation{Department of Physics, Stockholm University, S-10691,
 Stockholm, Sweden}


\date{\today}



\begin{abstract}
We report on an experimental test of classical and quantum dimension. We have used a dimension witness which can distinguish between quantum and classical systems of dimension 2,3 and 4 and performed the experiment for all five cases. The witness we have chosen is a base of semi-device independent cryptographic and randomness expansion protocols. Therefore, the part of the experiment, in which qubits were used, is a realization of these protocols.
In our work we also present an analytic method for finding the maximal quantum value of the witness along with corresponding measurements and preparations. This method is quite general and can be applied to any linear dimension witness.
\end{abstract}



\maketitle

Classical and quantum dimensions are fundamental quantities in information processing. In particular, security of many cryptographic schemes\cite{ACM06,PB11,LYWZWCGH11} crucially relies on the dimensional characteristics of the information carriers. The concept of a quantum dimension witness was first introduced for the dimension of the Hilbert space of composite systems tested locally \cite{BPAGMS08}. Later, a device-independent dimension witness was introduced in \cite{GBHA10} and the robustness of such witnesses was analyzed in \cite{D12}. More recently the device-independent dimension witnesses were realized experimentally \cite{ABCB12,HGMBAT12}.

Apart from testing the dimension of a system, the witnesses can also have a more practical application: semi device independent protocols. In these scenarios we do not make any assumptions on the devices that the parties involved are using but we do assume an upper bound on the dimension of the systems communicated. This setting provides a good compromise between fully device independent protocols and ones with complete knowledge of the devices because it makes implementation much easier than in the first case and provides better security than in the second. The notion of semi-device independence was introduced in \cite{PB11} in the context of cryptography and was later developed for randomness expansion in \cite{LYWZWCGH11,M-sdirng,M-divssdi} and for quantum state discrimination in \cite{BNV13}. These applications require witnesses based on quantum random access codes (QRAC) \cite{WCD08,Am1}. The witnesses realized in \cite{ABCB12,HGMBAT12} do not have this property.

In this work we analytically study and then experimentally realize a dimension witness inspired by the CHSH inequality \cite{CHSH69}. First we derive the bounds for the classical and quantum systems of dimensions 2,3 and 4 (the witness is saturated by a four level system and can not make distinction
for higher dimensions). Later we describe the experimental setup and present the results. Finally, we remark on how the test for quantum dimension 2, that we have conducted, would perform as a realization of a semi-device independent QKD or randomness expansion protocol.

\begin{figure}[h]
\centerline{\includegraphics[width=0.8\columnwidth]{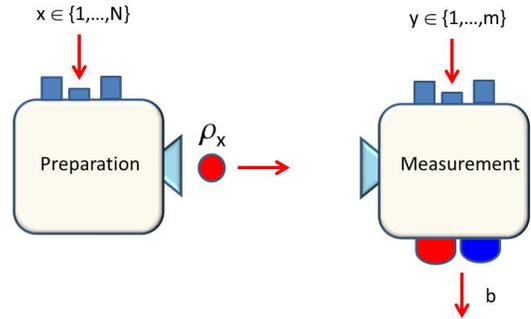}}
\caption{Device-independent scenario for testing the minimum classical or quantum dimension. When button $x\in\{1,\ldots,N\}$ is pressed, a state preparator part of the setup, with $N$ buttons, emits a particle in a state $\rho_x$. This particle is sent to a measurement device with $m$ buttons. When button $y\in\{1,\ldots,m\}$ is pressed, the device performs measurement $M_y$ on the particle. The measurement produces outcomes $b \in \{-1,+1\}$.}\label{Fig1}
\end{figure}

The scenario that we consider is schematically illustrated in Fig.~\ref{Fig1}. There is a state preparator with $N$ buttons; it emits a particle in a state $\rho_x$ (specified by the device's supplier) when button $x \in \{1, . . . ,N\}$ is pressed. For testing, the emitted particles are sent to a measurement device, with $m$ buttons. When button $y \in \{1, . . . ,m\}$ is pressed, the
device performs measurement $M_y$ on the incoming particle. The measurement produces outcome $b \in \{-1, +1\}$. A complete test should yield  probability distributions $P(b|x, y)$ for obtaining result $b$ in measurement $M_y$ on state $\rho_x$. Suitable combinations of the experimental probabilities $P(b|x, y)$ can then be compared with the theoretical classical and quantum bounds of the dimension witness.

Dimension of a system is defined as the number of distinguishable states. In classical information theory the states of a system are values of bits or trits and are all perfectly distinguishable. Therefore the classical dimension of a bit is $d_c=2$, of a trit $d_c=3$ and of a pair of bits $d_c=4$. In the quantum case the dimension $d_q$ is simply the dimension of the Hilbert space. For our tests of the lower bounds for $d_c$ and $d_q$ we utilize the dimension witnesses of the type introduced in \cite{GBHA10}. These witnesses use as primary quantities the expectation values
\begin{equation}
 \label{mean}
 E_{xy}= P(+1|x,y)-P(-1|x,y),
\end{equation}
In our test we use a CHSH-inspired combination of these expectation values which we denote $D_{CHSH} $. We call it CHSH-inspired because it can be obtained from the CHSH inequality using the method described in \cite{M-divssdi}. It involves four states ($N = 4$) and uses two dichotomic measurements ($m = 2$):
\beqa
 \label{ICHSH}
 D_{CHSH}  &\equiv& (E_{11} + E_{12})- (E_{21} + E_{22}) + \\ \nonumber
           &+&      (E_{31}- E_{32}) - (E_{41} - E_{42})
 \leq \lambda_d.
\eeqa

The upper bound $\lambda_d$ further on will be denoted as  $C_d$ and $Q_d$ for classical and quantum cases, respectively. The subscript $d$ denotes the dimension. Classical ensembles allow for statistical mixtures of identical or fully distinguishable states only.
Quantum ensembles permit pure states, which are neither identical nor orthogonal to each other. Since classical ensembles are more restricted than quantum, one immediately notices that $C_d \leq Q_d$.

To find the classical bounds $C_d$, notice that, due to linearity of $D_{CHSH}$, only deterministic strategies need to be considered. The preparator sends deterministic messages, but is constrained by the dimension of the  system.
Thus, a classical $d_c$ dimensional system can be linked with $d_c$ different two bit messages, each bit determining the system's behavior for a given  setting $y$ of the receiver. Each such a message can be put in a form of a two dimensional vector $\vec{v}^{x}$, with components $\vec{v}^{x}_y=\pm1$ giving the output of the receiver, for the given message/preparation $x$.  The classical deterministic value of  (\ref{ICHSH}) is
\be  \nonumber
D_{CHSH}=\vec{v}^{1}\cdot (1,1)+\vec{v}^{2}\cdot (-1,-1)
\\ \label{dchsh}
+\vec{v}^{3}\cdot (1,-1)+\vec{v}^{4}\cdot (-1,1).
\ee
If a vector $\vec{v}^{x}$ is equal to the vector, which enters the given scalar product with it,  their product is 2. If they differ in 1 component, this product is 0. If they differ in two the product is -2. Thus,  if all four $\vec{v}^{x}$ are different, and each is equal to the vector by which it is scalarly  multiplied in (\ref{dchsh}), the value of $D_{CHSH}$ can reach value 8,. Thus $d_c=4$ implies $C_4=8$. If there are only three different values of $\vec{v}^{x}$ then at most three of the terms in (\ref{dchsh}) can be 2.  Thus,  $C_3=6$. For $d_c=2$ we have $C_2=4$.

For the quantum bounds  the relevant measuring operators for $d = 2$ and $d = 3$ must obey the following:


{\it Lemma 1.} To find the maximal quantum value of any linear dimension witness, based on binary outcomes, given by
  \be W_D= \sum_{x=1}^N\sum_{y=1}^m\sum_{s=\pm1}K_{(x,y,s)}P(s|x,y),\ee
where $K_{(x,y,s)}$
are some real coefficients.  It is sufficient to consider only pure states and projective measurements.

{\em Proof.}
Since any mixed state is as a convex combination  (probabilistic mixture) of pure ones, the value of the part of a dimension witness corresponding to such a  state is equal to a probabilistic average of the values for  the pure states. Therefore, it cannot be greater than the largest value entering this sum. Thus the maximal value is achieved for pure states.  We only need to prove that projective measurements are sufficient.
The most general form of measurement is a positive operator value measurement (POVM). If there are only two outcomes, a POVM measurement consists of a pair of positive operators that sum up to identity. We denote them $\mathcal{O}^y_+$ and $\mathcal{O}^y_-=\openone-\mathcal{O}^y_+$. Obviously,  $\mathcal{O}^y_+$ and $\mathcal{O}^y_-$ commute. Thus, they can be simultaneously diagonalized. Therefore, we can write them as  $\mathcal{O}^y_-=\sum_{j=1}^dc^y_j|j\rangle\langle j|$ and  $\mathcal{O}^y_+=\sum_{j=1}^d(1-c^y_j)|j\rangle\langle j|$, where $|j\rangle$'s  form the diagonalizing basis ($d$ is the dimesion of the system). Obviously, $0\leq c_i \leq1$.  The probability of obtaining outcome $s=\pm 1$  when measuring system in pure state $\psi_x$ is $P(s|x,y)=\langle\psi_x| \mathcal{O}^y_s|\psi_x\rangle$, where $|\psi_x\rangle$ is the quantum state sent by the preparer.

The dimension witness is a sum of terms corresponding to different measurements. Such terms corresponding to a specific measurement setting $y$ are given by
\begin{eqnarray}
&\sum_x K(x,y,+1) \bra{\psi_x} \mathcal{O}_+^y \ket{\psi_x}&\nonumber
\\&
+\sum_x K(x,y,-1) \bra{\psi_x}\mathcal{O}_-^y  \ket{\psi_x}=k_0+\sum_j c_j k_j.&
\end{eqnarray}
The coefficients $k_0$ and
 $k_j$'s  can be easily calculated,  but their actual values are irrelevant. What is relevant, is the fact that the final formula is linear in $c_i$'s.
The maximal value of this expression is reached when $c_j$'s reach their boundary values, i.e., are equal  $1$  or $0$.  In all such cases $\mathcal{O}_{\pm}^y$ are projectors. QED.

Projective dichotomic measurements in for any dimension, of eigenvalues $\pm 1$, are represented by operators of the form
\begin{equation}
 \label{measurements}
 M_i = \eins - 2\ket{m_i}\bra{m_i},
\end{equation}
where $\eins$ denotes the identity matrix. In both  cases of dimensions two and three, and just two dichotomic operators ($i=1,2$), one can always find such a specific basis in which the  states linked with eigenvalues $-1$ can be put as
\begin{equation}
 \label{vectors}
 \ket{m_{1,2}} = \cos\left(\frac{\theta}{2}\right)\ket{1} \mp \sin\left(\frac{\theta}{2}\right)\ket{2}.
\end{equation}
In the case of qutrits this is so because $\ket{m_i}$ span a two dimensional sub-space, and one can define the basis state $\ket{0}$ as being orthogonal to it.
Moreover, the expectation value of $D_{CHSH}$ can be written as
\beqa
 \label{ICHSH2}
 & \bra{\psi_1}(M_1 + M_2)\ket{\psi_1} - \bra{\psi_2}(M_1 + M_2)\ket{\psi_2} &\nonumber \\
           &+     \bra{\psi_3}(M_1 - M_2)\ket{\psi_3} - \bra{\psi_4}(M_1 - M_2)\ket{\psi_4}.&
\eeqa
The optimization can thus be reduced to finding the maximum of the sum of the differences between the maximum and minimum  eigenvalues  of $M_1 + M_2$ and $M_1 - M_2$ with
\begin{equation}
 M_1 + M_2 = 2[\eins - (1 + \cos{\theta})|1\rangle \langle 1|- (1 - \cos{\theta})|2\rangle \langle 2|]
 \end{equation}
 and
 \begin{equation}
 M_1 - M_2 = 4\cos{(\frac{\theta}{2})}\sin{(\frac{\theta}{2})}[|1\rangle \langle 2| + |2\rangle \langle 1|]
 \end{equation}

 For $d = 2$, the Hilbert space is spanned by vectors $| 1\rangle$ and $| 2\rangle$. This gives
 \begin{equation}
 M_1 + M_2 = 2\cos{(\theta)}[|2\rangle \langle 2|- |1\rangle \langle 1|]
 \end{equation}
 and
 \begin{equation}
 M_1 - M_2 = 2\sin{(\theta)}[|1\rangle \langle 2| + |2\rangle \langle 1|]
 \end{equation}
 without loss of generality one can choose $\cos{(\theta)} \geq 0$ and $\sin{(\theta)} \geq 0$. This fixes the optimal states to
 \begin{subequations}
\begin{align}
 \ket{\psi_1} &= \ket{2},\\
 \ket{\psi_2} &= \ket{1}, \\
 \ket{\psi_3} &= \frac{1}{\sqrt{2}}(\ket{1} + \ket{2}), \\
 \ket{\psi_4} &= \frac{1}{\sqrt{2}}(\ket{1} - \ket{2}),
\end{align}
\end{subequations}
and reduces the optimization to finding the maximum value of $4(\cos{(\theta)} + \sin{(\theta)}) \leq 4\sqrt{2}$. The bound is achieved for $\theta = \pi/4$. Thus the qubit bound for $D_{CHSH} $ is $Q_2 = 4\sqrt{2} \sim 5.66$

For $d=3$, the Hilbert space is spanned by vectors $| 0 \rangle$ and $| 1 \rangle$ and $| 2 \rangle$,
 the sum $M_1 + M_2$ is
 \begin{equation}
 M_1 + M_2 = 2[|0\rangle \langle 0| + \cos{(\theta)}( |2\rangle \langle 2| - |1\rangle \langle 1|)].
 \end{equation}
Thus the optimal $\ket{\psi_1}$ becomes $\ket{0}$. Determination of the bound
of $D_{CHSH} $ for qutrits is thus reduced to the maximization of $(2 + 2\cos{\theta} + 4\sin{\theta}) \leq 2(1 + \sqrt{5})$, which is achieved for $\tan{(\theta)}=2$. Thus the qutrit bound for $D_{CHSH} $ is $Q_3 = 2(1 + \sqrt{5}) \sim 6.47$.

Note that due to our lemma if for example a qubit enter our device, then measurements of degenerate dichotomic observables of dimension larger than two constitute POVM measurements on a qubit (by Naimark theorem). Thus, in such a case the qubit limit in the inequality cannot be violated.

In general the dimension testing protocol, could be put as follows. A state preparator claims that his/her systems are of certain classical or quantum dimension, and the emitted systems are tested with observables selected in such a way that they are compatible with the claim. If
for example the claim is that the system is qutrit, and the bound for qubits is violated, then the system is of a higher dimension than two. If the value is close to the bound for qutrits we can safely conclude that the system has such a dimension as declared, and imperfections do not allow perfect saturation of the bound. Of course the system may be of even higher dimensionality, and in a more noisy state. Thus we effectively test the minimal dimension of the system provided by the preparator.


Let us now move to our experimental realization of the dimension indicator.

\begin{figure}[h]
\centerline{\includegraphics[width=0.8\columnwidth]{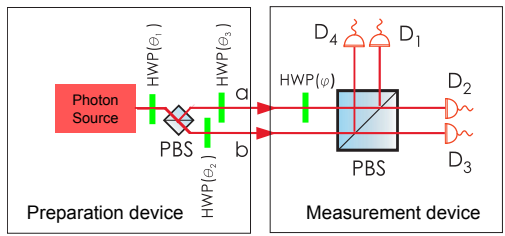}}
\caption{Experimental set-up for a witness  testing classical and
quantum dimensions.The state preparator is a single photon source emitting horizontally polarized photons which, after passing through three half-wave plates (HWP) suitably oriented at angles $\theta_i$ (with $i=1,2,3$), the photons are prepared in the required states. Information is encoded in horizontal and vertical polarizations, and in two spatial modes. The probabilities needed for the dimension witnesses $D_{CHSH}$ are obtained from the number of detections in the detectors $D_i$, after properly adjusting the orientation $\varphi$ of the half wave plate on the measurement side of the setup.  PBS stand for polarizing beam splitter.}
\label{Fig2}
\end{figure}

The preparer uses a preparation device (see the preparation device frame in Fig.~\ref{Fig2}),
which encodes
 the information in  four basis states: $|1\rangle \equiv |V,a\rangle$, $|2\rangle \equiv |H,a\rangle$, $|3\rangle \equiv |V,b\rangle$ and
$|0\rangle \equiv |H,b\rangle$, where   ($H$) and  ($V$) are horizontal and vertical polarization photonic modes respectively, and ($a$ and $b$) are two spatial photonic modes. Any qutrit state can be written as $\alpha |H,a\rangle + \beta |V,a\rangle + \gamma |H,b\rangle$, and any qubit state can be represented by $\alpha |H,a\rangle + \beta |V,a\rangle$.
The photonic states were prepared by three suitably oriented half-wave plates HWP($\theta_1$), HWP($\theta_2$) and HWP($\theta_3$) such that

\begin{equation}
 \begin{split}
 \ket{\psi} = &
 \sin{(2\theta_1)}\cos({2\theta_3)}|H,a\rangle -
 \sin{(2\theta_1)}\sin{(2\theta_3)}|V,a\rangle\\
 &+\cos{(2\theta_1)}\cos{(2\theta_2)}|H,b\rangle +
 \cos{(2\theta_1)}\sin{(2\theta_2)}|V,b\rangle.
 \end{split}
\end{equation}
Thus, by adjusting the orientation angles $\theta_i$ of the HWP$(\theta_i)$, we could produce any of the required states. In the experiment, classical sets (bits, trits, quarts) consisted of states which were perfectly distinguishable or identical.

The tester can use different operational approach, but with a single measurement device, depending on the claim of the preparer, and test whether the bounds are violated. To implement $M_i$, the HWP is set  in such a way that the eigenstate $\ket{m_i}$, which corresponds to the $-1$ eigenvalue, can give a click only in detector $D_1$.

The internal functioning of the measurement device is as follows: It consists of one adjustable HWP($\varphi$) and one polarization beam splitter (PBS).
If the input state is the eigenstate with negative eigenvalue, the polarization in mode $a$ will first be rotated by HWP($\varphi$) to obtain the state $\beta' \ket{V,a}$. Then the PBS splits the polarization modes of the two spatial modes, this will give a click  in detector $D_1$. All the tests are exactly the same, up to a half wave plate rotation. To test the qubit, $M_1$ and $M_2$ correspond to setting the half wave plate to an angle $\varphi = 11.25^\circ$ and $\varphi = 78.75^\circ $ respectively. To test the qutrit, $M_1$ and $M_2$ correspond to setting the half wave plate to an angle $\varphi = 15.86^\circ $ and $\varphi = 74.14^\circ $ respectively.

For qubit states, $P(+1|x,y)$ and $P(-1|x,y)$ were estimated from the number of detections in $D_2$ and  in $D_1$ respectively. For qutrit states, the values of $P(+1|x,y)$ and $P(-1|x,y)$ were obtained from the number of detections in $D_2$ and $D_3$, and in $D_1$, respectively. When the preparer claimed classical sets, the measurement settings of the tester were reduced to arranging the detectors so that they clicked with the negative eigenvalue upon receiving a photon in a particular basis state: $|0\rangle \rightarrow D_3$, $|1\rangle \rightarrow D_1$, $|2\rangle \rightarrow D_2$ and $|3\rangle \rightarrow D_4$.

Our single-photon source was weak coherent light from a  diode laser emitting at 780 nm. The laser was attenuated so that the two-photon coincidences were negligible. Our single-photon detectors ($D_i$, $i=1,2,3,4$) were Silicon avalanche photodiodes with detection efficiency $\eta_d = 0.55$ and dark counts rate $R_d \simeq 400$ Hz.

The detectors $D_i$ produced output transistor-transistor logic signals of 4.1 V (with duration of $41$ ns). The dead time of the detectors was $50$ ns. All single counts were registered using multi-channel coincidence logic with a time window of $1.7$ ns.




The goals of the experiments were to obtain the maximum qubit violation of the bit bound $D_{CHSH} (\text{bit})=C_2 = 4$, the maximum trit violation of the qubit bound $D_{CHSH}  (\text{qubit})= Q_2 = 5.66$, the maximum qutrit violation of the trit bound $D_{CHSH}  (\text{trit}) = C_3 = 6$. We prepared four qutrit states and performed $m=2$ dichotomic measurements which maximize $D_{CHSH} $. The last experiment was a $D_{CHSH} $ test on quarts, the maximum quart violation of the qutrit bound $D_{CHSH}  (\text{qutrit}) =Q_3 = 6.47$. For this, we prepared four fully distiguishable  quart states the $D_{CHSH} $, and the results were  very close to the algebraic bound $D_{CHSH} = C_4 = 8$.

All our experimental results are summarized in Tab.~\ref{tab}. The  experimental values are in a very good agreement with the theoretical predictions. This clearly demonstrates that we are able to determine the minimum dimension of a supplied set of states. The errors were deduced from propagated Poissonian counting statistics of the raw detection events, the limited precision of the settings of the polarization components (HWP plates) and the imperfection of the polarizing beam splitters. The number of detected single photons was about $1.5 \times 10^5$ per second and the coincidences to singles ratio was less than $2 \times 10^{-4}$. The measurement time for each experiment was $30$ s. All the results and their corresponding errors are listed in Table 1.

\begin{table}[h]
\begin{center}
\begin{tabular}{lllllll}
 Inequality bound     & $D_{th}$ & $D_{exp}$ & $D_{exp}^b$ & $\Delta D_{p}$ & $\Delta D_{d}$ & $\Delta D_T$ \\
 \hline
 $D_{CHSH} $(bit)      & 4     & 3.94  & 3.98 & 0.08  & 0.010 & 0.08 \\
 $D_{CHSH} $(qubit)    & 5.66  & 5.51  & 5.56 & 0.12  & 0.008 & 0.12 \\
 $D_{CHSH} $(trit)     & 6     & 5.90  & 5.96 & 0.13  & 0.010 & 0.13 \\
 $D_{CHSH} $(qutrit)   & 6.47  & 6.44  & 6.50 & 0.14  & 0.009 & 0.14 \\
 $D_{CHSH} $(quart)    & 8     & 7.88  & 7.94 & 0.16  & 0.010 & 0.16 \\
\end{tabular}
\end{center}
\caption{{\bf Experimental results of the dimension witness tests.}
$D_{th}$, $D_{exp}$ and $D_{exp}^b$ represent the theoretical, raw
experimental and dark counts corrected experimental values of the
dimension witness bounds, respectively. $\Delta D_{p}$, $\Delta
D_{d}$ and $\Delta D_T$ are the errors due to the limited
precision of the settings of the polarization components and the
imperfections of the polarization splitting, the propagated
Poissonian counting statistics of the raw detection events and the
total errors, respectively.} \label{tab}
\end{table}

As we have previously stated, the witness $D_{CHSH}$ plays a crucial role in semi-device independent quantum key distribution and randomness expansion. In \cite{PB11} it has been used as a certificate for the security of quantum key distribution. The protocol there assumed that the communicated system was a qubit and this has been the only assumption on the devices made. The value of $D_{CHSH}$ for qubits obtained in our experiment would imply a secure key rate of 5.18\% or 6.67\% if we had made an additional assumption that the dark counts observed are not controlled by a potential eavesdropper. The derivation of these values is given in the appendix.

The witness $D_{CHSH}$ has been first used as a certificate for semi-device randomness expansion in \cite{LYWZWCGH11} and the bounds on the amount of min-entopy generated with it have been improved in \cite{M-divssdi}. The amount of min-entropy generated by a round of protocol with qubits for the values observed in our experiment can be read from fig. 4. in \cite{M-divssdi} (note that $T_2$ used there is equal to $\frac{1}{2}D_{CHSH}$). For $D_{CHSH}=5.51$ it is 0.0595 and for $D_{CHSH}=5.56$ it reaches 0.0820.




In summary: We have experimentally determined lower bounds for the dimension of several ensembles of physical systems in a device-independent way. For the tests we utilized a dimension witnesses device inspired by the structure of the CHSH inequality which has a direct application in semi-device independent quantum key distribution and randomness expansion. Note that the presented single device is universal for all studied dimensions from $d_{c/q}=2$ to $d_{c/q}=4$. In the  witness device we used optimal measurements for the given dimension. We applied them to sets of photonic bits,
qubits, trits, qutrits, and quarts. Our results demonstrate that CHSH inspired dimension witnesses can be utilized to
test classical and quantum dimensions of sets of physical states
generated in externally supplied, potentially defective devices, and
that one can distinguish between classical and quantum sets of states
of a given dimension. We have also discussed how efficient the semi-device independent protocols based on our witness would be with the values of it that we have reached in our experiment. The approach can be generalized to tests of
systems of higher dimensions. This can be done in several ways, some of which will be presented in forthcoming papers.


MP would like to thank Ryszard Weinar for discussions and Piotr Mironowicz for supplying the raw data from \cite{M-divssdi}. JA, PB and MB are supported by the Swedish Research Council (VR), the Linnaeus Center of Excellence ADOPT. MP is supported by TEAM program of foundation for Polish Science (FNP), UK EPSRC, NCN grant 2013/08/M/ST2/00626 and QUASAR (ERA-NET CHIST-ERA 7FP UE). MZ is supported by an  Ideas Plus Program MNiSW (IdP2011 000361).



\begin{thebibliography}{20}
\itemsep=0mm
\bibitem{ACM06}
 A. Ac\'{\i}n,  N. Gisin, and  L. Masanes,
 Phys. Rev. Lett. \textbf{97} 120405 (2006).

\bibitem{PB11}
 M. Paw{\l}owski and N. Brunner,
Phys. Rev. A \textbf{84}, 010302 (2011).

\bibitem{LYWZWCGH11}
 H. -W. Li, \emph{et al.}
 Phys. Rev. A \textbf{84}, 034301 (2011).

\bibitem{BPAGMS08}
N.  Brunner, \emph{et al.}
Phys. Rev. Lett. \textbf{100}, 210503 (2008).

\bibitem{GBHA10}
 R. Gallego, N. Brunner, C. Hadley, and A.  Ac\'{\i}n,
 Phys. Rev. Lett. \textbf{105}, 230501 (2010).

\bibitem{D12}M. Dall�Arno, \emph{et al.} Phys. Rev. A \textbf{86}, 042312 (2012).


\bibitem{ABCB12}J. Ahrens, P. Badziag, A. Cabello, and  M. Bourennane, Nature Phys. 	\textbf{8}, 592 (2012).
\bibitem{HGMBAT12}M. Hendrych,	R. Gallego,	M. Micuda, N. Brunner,	A. Ac\'{\i}n,	and  J. P. Torres	
 Nature Phys. 	\textbf{8}, 588 (2012).

\bibitem{M-sdirng} H-W. Li, M. Paw{\l}owski, Z-Q. Yin, G-C. Guo, and Z-F. Han, Phys. Rev. A {\bf 85},  052308 (2012).
\bibitem{M-divssdi} H.-W. Li, \emph{et al.} Phys. Rev. A {\bf 87}, 020302(R) (2013).

\bibitem{BNV13} N. Brunner, M. Navasques, and T. V�rtesi, Phys. Rev. Lett. \textbf{110}, 150501 (2013).

\bibitem{WCD08}S. Wehner, M. Christandl, and A. C. Doherty, Phys. Rev. A \textbf{78}, 062112 (2008).

    \bibitem{Am1} A. Ambainis, A. Nayak, A. Ta-Shma, U. Vazirani, Journal of the ACM, {\bf 49(4)} 496 (2002).

\bibitem{CHSH69} J.F. Clauser, M.A. Horne, A. Shimony, and R.A. Holt, Phys. Rev. Lett. \textbf{23}, 880 (1969).
\bibitem{CK} I. Csiszar and J. K\"orner, IEEE Trans. Inf. Theory {\bf 24}, 339 (1978).

\bibitem{M-hyper} M. Paw{\l}owski, A. Winter, Phys. Rev. A {\bf 85}, 022331 (2012).
 \end{thebibliography}
\end{document}